\documentclass{article}
\usepackage[utf8]{inputenc}
\usepackage{authblk}
\usepackage{indentfirst}
\usepackage{hyperref}
\usepackage{longtable}
\usepackage{setspace}
\usepackage[margin=1in]{geometry}
\usepackage{graphicx}
\graphicspath{ {./Figures/} }
\usepackage{subcaption}
\usepackage{amsmath}
\usepackage{lineno}
\usepackage{multirow}
\usepackage{booktabs}
\usepackage{siunitx}
\usepackage{array}
\usepackage{multirow}
\usepackage{color,soul}
\usepackage[version=4]{mhchem}
\usepackage{float} 

\usepackage[biblabel]{cite}

\title{First-principles investigation of Sr-Ce-M-O perovskites for solar thermochemical water splitting}

\author[1]{Sachin Kumar}
\author[1]{Pritam Ghosh}
\author[1,*]{Gopalakrishnan Sai Gautam}

\affil[1]{Department of Materials Engineering, Indian Institute of Science, Bengaluru, 560012, India}
\affil[*]{Email: \href{mailto:saigautamg@iisc.ac.in}{saigautamg@iisc.ac.in}}

\begin{document}

\maketitle

\begin{abstract}
Using density functional theory based calculations, we systematically examine the utility of Sr-M-O and Sr-Ce-M-O perovskites for solar thermochemical water splitting, a promising route for sustainable hydrogen production. Importantly, we identify Sr$_{0.5}$Ce$_{0.5}$MnO$_3$ and Sr$_{0.5}$Ce$_{0.5}$CrO$_3$ to be promising candidates, exhibiting optimal oxygen vacancy formation energy and 0 K thermodynamic stability.
\end{abstract}


\section{Introduction}
The escalating global energy demand and the pressing need to mitigate climate change necessitate a transition towards sustainable and carbon-neutral energy carriers.\cite{IEA2023,Armaroli2007} Hydrogen, with its high energy density and zero-emission combustion, stands out as a promising alternative to fossil fuels.\cite{Turner1999,Falcone2021} To produce green hydrogen on a large scale, solar thermochemical water splitting (STWS) cycles offer a direct and efficient pathway by harnessing concentrated solar energy to drive the endothermic dissociation of water, typically via a two-step redox cycle involving a metal oxide (see Section~I of the electronic supplementary information$-$ESI).\cite{Steinfeld2005,Agrafiotis2015}

The STWS cycle generally comprises:
\begin{enumerate}
    \item \textbf{Thermal Reduction (TR):} MO\(_x\) \(\rightarrow\) MO\(_{x-\delta}\) + \(\delta/2\) O\(_2\)(g) [at $T >$1673~K]
    \item \textbf{Water Splitting (WS):} MO\(_{x-\delta}\) + \(\delta\) H\(_2\)O(g) \(\rightarrow\) MO\(_x\) + \(\delta\) H\(_2\)(g) [at $T$ between 873-1473~K]
\end{enumerate}

Non-stoichiometric cerium dioxide (CeO\(_2\)) is the current state-of-the-art material for STWS owing to its rapid redox kinetics and structural stability.\cite{Chueh2009,muhich2017principles} However, CeO\(_2\) requires very high reduction temperatures (>\,1500\,$^\circ$C) to achieve even modest extent of reduction (i.e., \(\delta\)), limiting practical solar-to-hydrogen efficiency.\cite{Bulfin2017} Specifically, CeO$_2$ exhibits an enthalpy of reduction ($\Delta H_{\text{red}}$)$\sim$4~eV per oxygen vacancy, which is significantly higher than the ideal range of 3.4-3.9~eV/vacancy for optimal STWS operating conditions.\cite{wexler2023materials} Offering tunable redox properties, perovskite oxides (ABO\(_3\)) have emerged as attractive alternatives to CeO$_2$ for STWS, exhibiting favourable thermodynamics for reduction at moderate temperatures (between 1200–1400\,$^\circ$C) and good kinetic performance.\cite{McDaniel2013,Vieten2016,witman2023defect} Most perovskites, however, exhibit redox activity primarily on the B-site transition-metal cation, restricting the configurational entropy change upon reduction (\(\Delta S_{\text{red}}\)) relative to ceria,\cite{naghavi2017giant} eventually limiting amount of H$_2$ generated.

At any $T$, the Gibbs energy of reduction (\(\Delta G_{\text{red}}=\Delta H_{\text{red}}-T\Delta S_{\text{red}}\)) governs the extent of reduction. For a given \(\Delta H_{\text{red}}\), a higher \(\Delta S_{\text{red}}\) permits greater reduction extent at a given temperature, or, equivalently, greater reduction at lower temperatures. Consequently, designing perovskites with enhanced \(\Delta S_{\text{red}}\) by enabling multi-site redox is a key strategy for improving STWS performance. Previous work on Ca\(_{0.5}\)Ce\(_{0.5}\)MnO\(_3\) demonstrated the feasibility of simultaneous A-site (Ce) and B-site (Mn) redox, leading to improved STWS characteristics.\cite{Gautam2020,wexler2023multiple} Building upon this concept, Sr-based perovskites are of particular interest owing to Sr’s larger ionic radius, which may accommodate redox-active Ce more readily\cite{gao2023remarkable,heo2021double} and also influence oxygen mobility. Nevertheless, a systematic investigation of Sr-based quaternary perovskites incorporating Ce for STWS is currently lacking.

Our study employs density functional theory (DFT) based calculations to systematically explore the potential of Sr-M-O ternary (SrMO\(_3\)) and Sr-Ce-M-O quaternary (Sr\(_{0.5}\)Ce\(_{0.5}\)MO\(_3\)) perovskites (where M = Ti, V, Cr, Mn, Fe, or Co) as potential STWS materials. We exclude Sc and Zn from the 3$d$ series since they are not redox-active, while we exclude Cu since it typically does not acquire a +3 oxidation state and form perovskite structures.\cite{krzystowczyk2020substituted} We also exclude Ni since we could not obtain a reliable experimental SrNiO$_3$ structure, which can be attributed to its instability under ambient pressures.\cite{takeda1972synthesis} We focus on elucidating the oxygen vacancy formation energy (E[V\(_{\text{O}}\)]), a proxy for \(\Delta H_{\text{red}}\), and 0~K thermodynamic stability to identify candidates potentially capable of simultaneous A-site (Ce\(^{4+}\)) and B-site (M-cation) reduction. Such dual-site redox activity can achieve higher \(\Delta S_{\text{red}}\),\cite{sai2020first} potentially surpassing CeO\(_2\) in STWS efficiency. Our findings identify Sr\(_{0.5}\)Ce\(_{0.5}\)CrO\(_3\) and Sr\(_{0.5}\)Ce\(_{0.5}\)MnO\(_3\) systems to be promising candidates, exhibiting optimal E[V\(_{\text{O}}\)] and thermodynamic stability, paving the way for a new class of high-performance STWS materials. 

\newpage
\section{Methods}
All spin-polarized DFT calculations were performed using the Vienna ab initio simulation package (VASP).\cite{Kresse1996PRB, kresse_ab_1993} We employed the projector augmented-wave potentials,\cite{Kresse1999PRB_PAW} and described the exchange-correlation interactions using the strongly constrained and appropriately normed (SCAN) functional.\cite{Sun2015PRL} A Hubbard $U$ correction (i.e., SCAN+$U$) was applied to the 3$d$ orbitals of M and the 4$f$ orbitals of Ce to accurately describe the strongly correlated electrons, with the $U$ values taken from literature.\cite{sai2018evaluating, long2020evaluating, swathilakshmi2023performance}We used a plane-wave energy cutoff of 520~eV and sampled the irreducible Brillouin zone using \(\Gamma\)-centered Monkhorst-Pack\cite{monkhorst1976special} $k$-point meshes with a density of at least 48 Å\(^{-1}\). We relaxed all structures, i.e., allowed cell shape, volume, and ionic positions to change without preserving symmetry, until atomic forces were below |0.01| eV/Å and the total energy converged to \(10^{-5}\) eV.

Initial structures for ternary SrMO\(_3\) perovskites were obtained from the inorganic crystal structure database (ICSD).\cite{Hellenbrandt2004CrystallogrRev} For each SrMO\(_3\) system, we considered various known polymorphs, as available in the ICSD, to identify the ground state structure (see Table~S1 of the ESI). Quaternary Sr\(_{0.5}\)Ce\(_{0.5}\)MO\(_3\) structures were generated by substituting 50\% of Sr atoms with Ce in the A-site of the ground state, ternary SrMO\(_3\) supercells. We utilized the pymatgen library\cite{Ong2013ComputMaterSci} to generate symmetrically distinct A-site cation orderings (i.e., Sr and Ce configurations) within each SrMO$_3$ unit cell or supercell. 

The thermodynamic stability of pristine perovskites was evaluated by calculating the energy above the convex hull (E\(_{\text{hull}}\)) with respect to other competing phases. For calculating E$_{\text{hull}}$, we considered all ordered unary, binary, ternary, and quaternary Sr-Ce-M-O compounds, as available in the ICSD, excluding intermetallics, as compiled in Section~V of the ESI. While E\(_{\text{hull}}\) > 0 indicates instability (or metastability) at 0~K, entropic contributions at high operational temperatures can stabilize some metastable phases to an extent.\cite{Gautam2020} Convex hull plots of ternary and quaternary systems are compiled in Figures~S1 and S2 of the ESI.

We calculated neutral E[V$_\text{O}$] within supercells of the pristine ground-state SrMO$_3$ and Sr$_{0.5}$Ce$_{0.5}$MO$_3$ structures. We maintained a minimum distance of 8~{\AA} between periodic vacancy images to reduce spurious interactions between the vacancy defect and its periodic images.\cite{Gautam2020} Specifically, we used 3$\times$3$\times$3 supercells for cubic and tetragonal systems, 2$\times$2$\times$2 supercells for orthorhombic and monoclinic systems and 2$\times$2$\times$1 supercells for hexagonal systems, respectively, to ensure convergence of E[V\(_{\text{O}}\)] to within $\pm$0.1 eV.\cite{wexler_factors_2021} We used the following equation to calculate E[V\(_{\text{O}}\)]:
\begin{equation} \label{eq:Evac}
\text{E}[\text{V}_{\text{O}}] = E_{\text{defective}}^{\text{SCAN+U}} - E_{\text{pristine}}^{\text{SCAN+U}} + \frac{1}{2} E_{\text{O}_2\text{(g)}}^{\text{SCAN}}
\end{equation}
where \(E_{\text{defective}}^{\text{SCAN+$U$}}\) and \(E_{\text{pristine}}^{\text{SCAN+$U$}}\) represent the total energy of the supercell with one oxygen vacancy and the pristine supercell, respectively, calculated with SCAN+$U$. \(E_{\text{O}_2\text{(g)}}^{\text{SCAN}}\) is the total energy of an isolated oxygen molecule in its triplet ground state, calculated using the SCAN functional. We considered all symmetrically distinct oxygen sites for the formation of the vacancy. For defective structures, we relaxed the ionic positions only with fixed lattice parameters (from the corresponding relaxed pristine ground state structure). We prioritized the lowest E[V\(_{\text{O}}\)] among different vacancy sites for identifying candidate compositions.

\newpage
\section{Results and Discussion}
\subsection{Ternary perovskites}
\begin{figure}[h!]
    \centering
    \includegraphics[width=\linewidth]{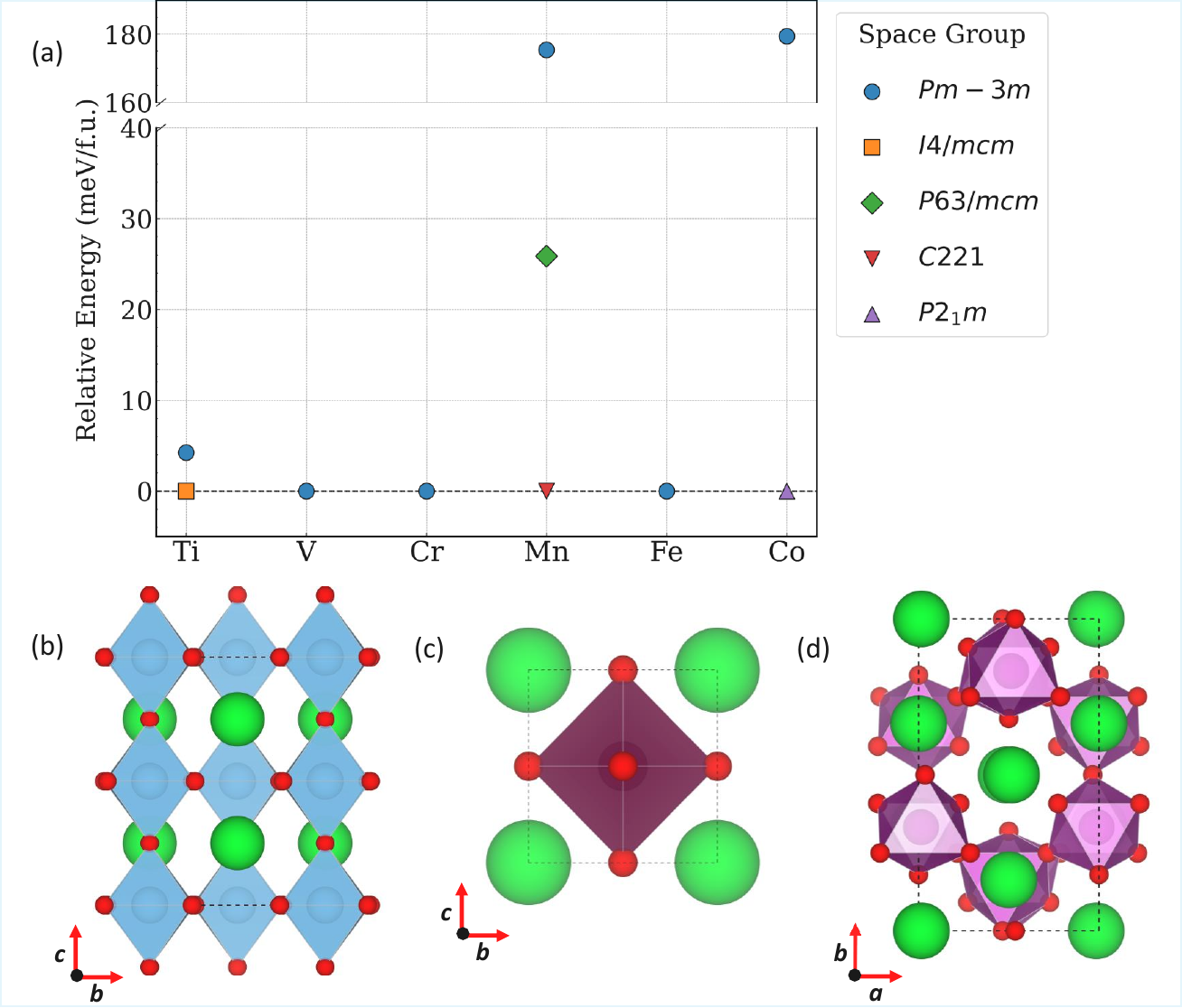}
    \caption{(a) Energy relative to the calculated ground state for different SrMO$_3$ polymorphs considered. (b-d) Ground state polymorphs of SrMO$_3$ ternary perovskites, namely (b) SrTiO$_3$, (c) SrVO$_3$, SrCrO$_3$, SrFeO$_3$, and SrCoO$_3$, and (d) SrMnO$_3$. Green and red spheres are Sr and O, respectively, while colored polyhedra represent MO$_6$ octahedra.}
    \label{fig:ternary_polymorphs}
\end{figure}

First, we use DFT to identify the ground state structures of ternary SrMO\(_3\) perovskites, among the various polymorphs reported in the ICSD, with Figure~{\ref{fig:ternary_polymorphs}}a plotting the energies (in meV/f.u.) relative to the calculated ground state for each SrMO$_3$. Specifically, the calculated ground states are tetragonal (I4/mcm) for SrTiO$_3$ (panel b in Figure~\ref{fig:ternary_polymorphs}), cubic (Pm$\overline{3}$m) for SrVO$_3$, SrCrO$_3$, SrFeO$_3$, and SrCoO$_3$ (panel c), and orthorhombic (C221) for SrMnO$_3$ (panel d), largely consistent with available experimental literature.\cite{Lan2003,Zhu2012,Wang2001,Bezdicka1993}  

In the case of SrMnO$_3$, we predict the orthorhombic structure to be more stable than the experimentally observed room temperature hexagonal (P6\(_3\)/mmc) phase,\cite{dong2023structure} which can be attributed to possible subtle low-temperature distortions that may be challenging to capture experimentally. In any case, we don't expect the orthorhombic SrMnO$_3$ phase to be relevant under STWS operating temperatures and utilize the hexagonal structure for subsequent calculations. In terms of the 0~K thermodynamic stability, we find all the calculated ground states of all SrMO$_3$ perovskites considered to be on the convex hull (i.e., E$_\text{hull} =$0~meV/atom), except SrCoO$_3$ (E$_\text{hull} \sim$200~meV/atom) and SrCrO$_3$ (E$_\text{hull} \sim$30~meV/atom, see Figure~S2 of the ESI). The hexagonal-SrMnO$_3$ is above the 0~K Sr-Mn-O convex hull by $\sim$5~meV/atom.

\begin{figure}[h!]
     \centering
     \includegraphics[width=\linewidth]{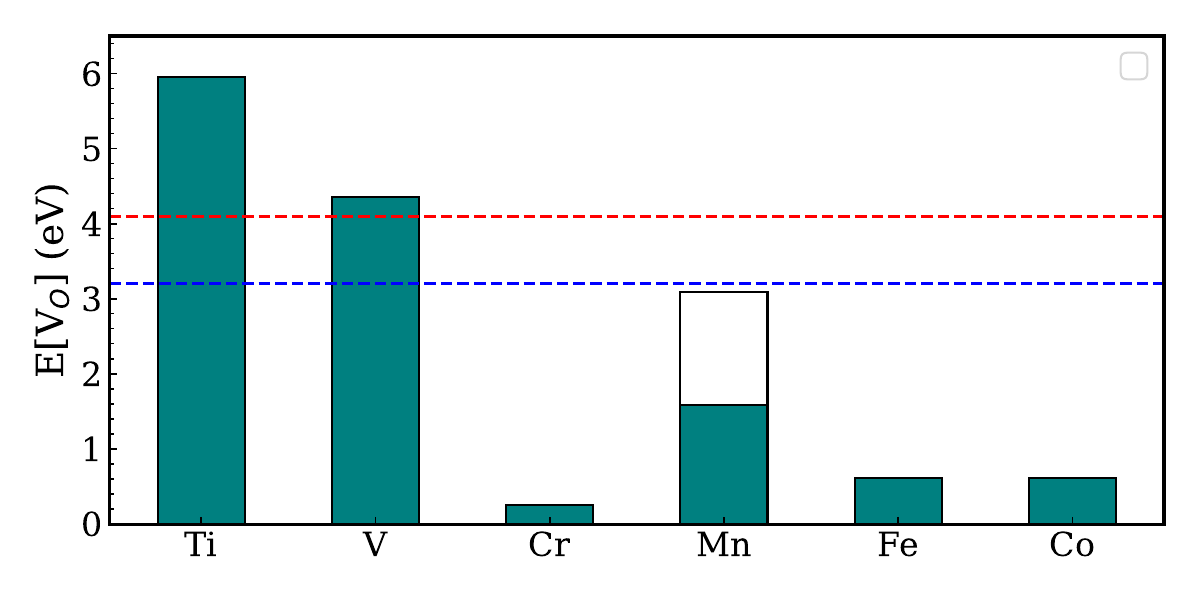}
     \caption{Calculated E[V$_{\text{O}}$] of the various ground state polymorphs of Sr-M-O ternary perovskites. The blue and red dashed line indicates the optimal range of  window of E[V$_{\text{O}}$] for STWS. For SrMnO$_3$, we considered the experimentally-relevant hexagonal polymorph instead of the ground state orthorhombic structure.}
     \label{fig:ternary_Evac}
\end{figure}

The calculated E[V\(_{\text{O}}\)] for the ternary Sr-M-O perovskites are shown in Figure~\ref{fig:ternary_Evac} for each M. While the solid bars indicate the lowest (or only) E[V\(_{\text{O}}\)] for each system, hollow bars signify the variation in E[V\(_{\text{O}}\)] due to differences in the oxygen vacancy configuration within a structure. The dashed blue and red lines indicate the optimal range of E[V\(_{\text{O}}\)] used for identifying potential STWS candidates, namely 3.2-4.1~eV, as used in prior literature.\cite{Gautam2020,wexler2023multiple,wexler2023materials} We require an optimal E[V\(_{\text{O}}\)] since the interactions between the oxide ion and its cation neighbors should be neither too strong (affects thermal reduction) nor too weak (affects spontaneity of water splitting).\cite{Gautam2020} 

We find the E[V\(_{\text{O}}\)] to follow the trends in the crystal reduction potentials of the M$^{4+}$, suggesting that the ease/difficulty of reducing the transition metal cation plays a crucial role in setting the magnitude of E[V\(_{\text{O}}\)] in ternary SrMO$_3$.\cite{wexler_factors_2021} For example, SrTiO$_3$ exhibits a high E[V\(_{\text{O}}\)] ($\sim$6~eV, Figure~{\ref{fig:ternary_Evac}}), well beyond the optimal 3.2-4.1~eV range, which can be attributed to the low (negative) crystal reduction potential of Ti$^{4+}$,\cite{wexler_factors_2021} making it too difficult to reduce under practical STWS conditions. On the other hand, SrCrO$_3$, SrFeO$_3$, and SrCoO$_3$ exhibit low E[V\(_{\text{O}}\)] ($\leq$2~eV), indicating insufficient ability to generate H$_2$. Importantly, we find SrVO$_3$ (E[V\(_{\text{O}}\)] \(\sim\) 4.35~eV) and hexagonal-SrMnO$_3$ (E[V\(_{\text{O}}\)]\(\sim\)1.59-3.09~eV) to be promising candidates, with the calculated E[V\(_{\text{O}}\)] close to the targeted range. However, the redox-activity of both SrVO$_3$ and SrMnO$_3$ is on the M-site only (see Table~S2 of the ESI).

\subsection{Quaternary perovskites}
\begin{figure}[h!]
    \centering
    \includegraphics[width=\linewidth]{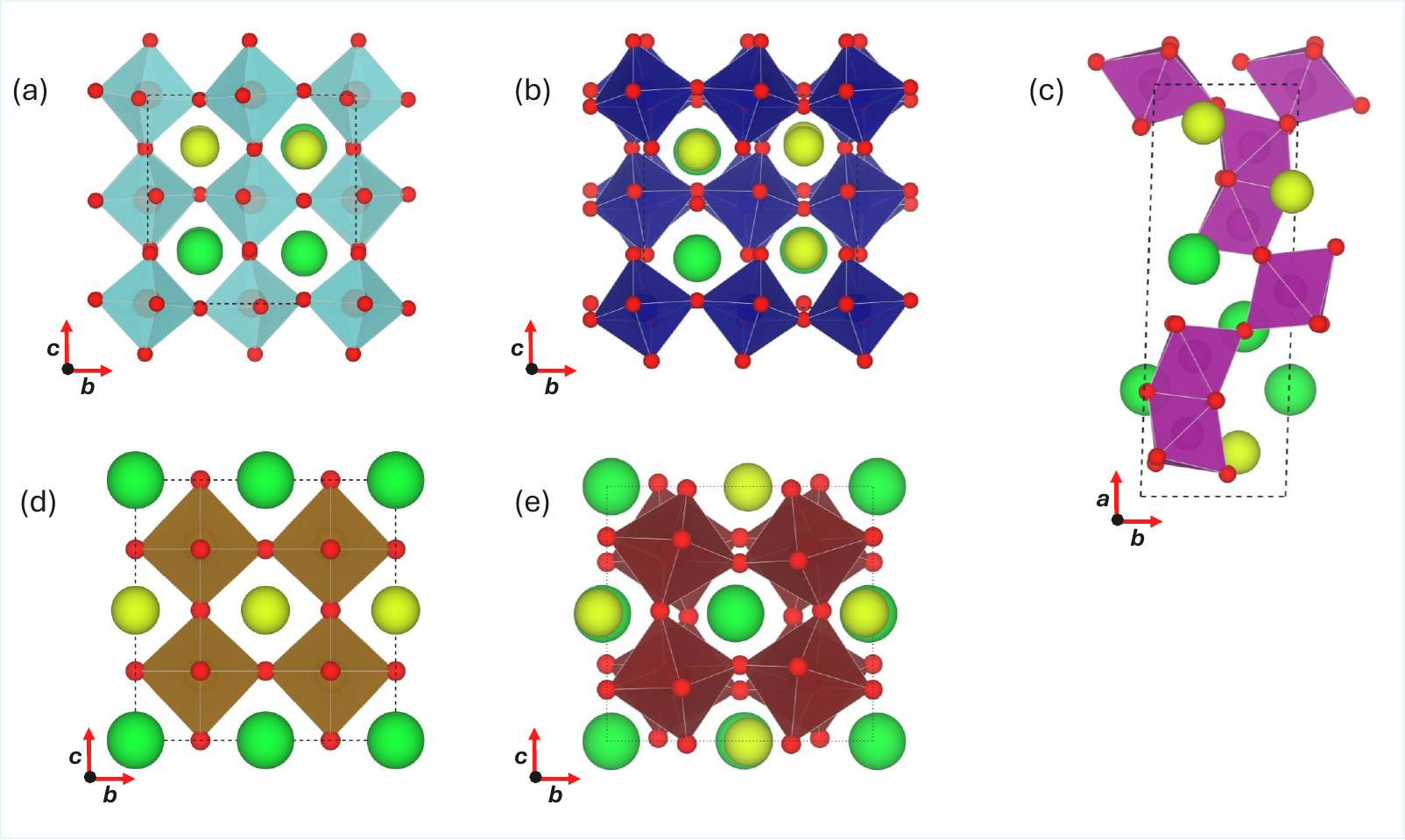}
    \caption{Ground state polymorphs of Sr\(_{0.5}\)Ce\(_{0.5}\)MO\(_3\) based perovskites, with notations similar to Figure~{\ref{fig:ternary_polymorphs}} and Ce atoms depicted by yellow spheres. (a) Sr\(_{0.5}\)Ce\(_{0.5}\)VO\(_3\), (b) Sr\(_{0.5}\)Ce\(_{0.5}\)CrO\(_3\), (c) Sr\(_{0.5}\)Ce\(_{0.5}\)MnO\(_3\), (d) Sr\(_{0.5}\)Ce\(_{0.5}\)FeO\(_3\), and (e) Sr\(_{0.5}\)Ce\(_{0.5}\)CoO\(_3\).}
    \label{fig:quaternary_structures}
\end{figure}

To potentially enhance \(\Delta S_{\text{red}}\), we introduce Ce to the A-site of the SrMO$_3$ ternaries  and create Sr\(_{0.5}\)Ce\(_{0.5}\)MO\(_3\) quaternaries. The potential presence of Ce\(^{4+}\), which can reduce to Ce\(^{3+}\), offers the possibility of redox activity at the A site in addition to the B site. Due to the high E[V\(_{\text{O}}\)] in ternary-SrTiO\(_3\), we did not consider the corresponding quaternary, i.e., Sr$_{0.5}$Ce$_{0.5}$TiO$_3$ for further calculations. Upon substitution of Sr with Ce, we used DFT to identify the ground state Sr-Ce configuration within each perovskite, as compiled in Figure~{\ref{fig:quaternary_structures}}. Notably, we observe all the quaternary ground state structures to be thermodynamically stable, i.e., E$_{\text{hull}} =$0~meV/atom (Figure~S2 of the ESI), indicating their potential synthesizability in experiments.

\begin{figure}[h!]
     \centering
     \includegraphics[width=\linewidth]{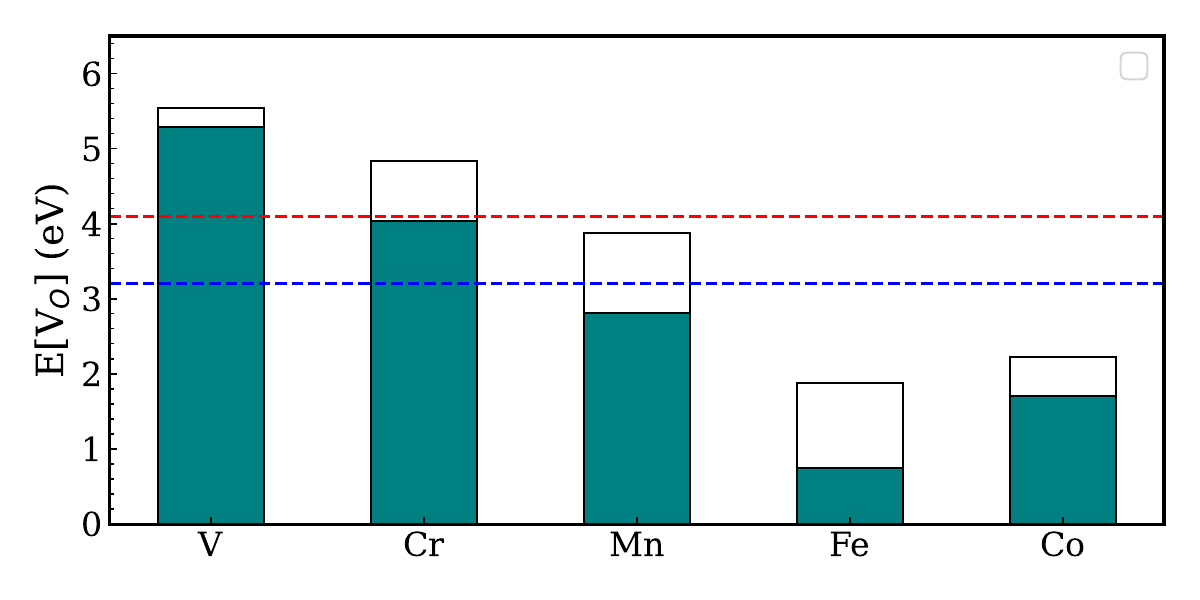}
     \caption{Oxygen vacancy formation energies of the various ground state polymorphs of Sr and Ce based quarternary perovskites. The blue and red dashed line indicates the suitable window of oxygen vacancy formation energies for STC application. Sr\(_{0.5}\)Ce\(_{0.5}\)CrO\(_3\) and Sr\(_{0.5}\)Ce\(_{0.5}\)MnO\(_3\) have values closer to the optimal range and are suitable candidates. }
     \label{fig:quat_evac}
\end{figure}

Figure~{\ref{fig:quat_evac}} displays the DFT-calculated E[V\(_{\text{O}}\)] for the Sr-Ce-M-O quaternaries considered. The trends in the calculated E[V\(_{\text{O}}\)] among the quaternaries do exhibit significant differences compared to the trends in ternary Sr-M-O, which can be due to the electrostatic and possible redox contributions of Ce. For example, the calculated E[V\(_{\text{O}}\)] do not differ by $\sim$4~eV between the Sr-Ce-V-O and Sr-Ce-Cr-O quaternaries, unlike the Sr-V-O and Sr-Cr-O ternaries. 

Notably, Sr\(_{0.5}\)Ce\(_{0.5}\)VO\(_3\) shows E[V\(_{\text{O}}\)] between 5.28$-$5.54~eV, which is too high for efficient STWS, despite the potential for V-site redox in ternary Sr-V-O. Also, Sr\(_{0.5}\)Ce\(_{0.5}\)FeO\(_3\) and Sr\(_{0.5}\)Ce\(_{0.5}\)CoO\(_3\) exhibit low E[V\(_{\text{O}}\)] ($\leq$2.2~eV), making them unsuitable, similar to their ternary counterparts. Importantly, we find Sr\(_{0.5}\)Ce\(_{0.5}\)MnO\(_3\) and Sr\(_{0.5}\)Ce\(_{0.5}\)CrO\(_3\) to be promising candidates for SWTS, given their calculated E[V\(_{\text{O}}\)], namely 2.81-3.87~eV and 4.03-4.83~eV, overlap with our target range of 3.2-4.1~eV. Note that Mn is a recurring redox-active element of importance among perovskites for STWS, since Mn-containing compositions have been demonstrated before as promising STWS candidates.\cite{Gautam2020,wexler2023multiple,barcellos2018}

Although we do not observe any dual cation redox at 0~K in any of the Sr-Ce-M-O quaternaries (see Table~S3 of the ESI), as indicated by changes in on-site magnetic moments, we expect multiple redox mechanisms to be activated at higher temperatures as vacancy concentration increases and higher energy vacancy configurations form. Simultaneous redox activity can also be induced by doping elements in addition to Ce on the B-site and/or on the A-site to form quinary and higher component Sr-perovskites.

\section{Conclusion}
In conclusion, we systematically investigated ternary SrMO\(_3\) and quaternary Sr\(_{0.5}\)Ce\(_{0.5}\)MO\(_3\) (M = Ti, V, Cr, Mn, Fe, or Co) perovskites for STWS using DFT-based calculations. We identified the ground state structures of both ternaries and quaternaries, evaluated their 0~K thermodynamic stability and calculated E[V$_{\text{O}}$]. Importantly, we found SrMnO$_3$ and SrVO$_3$ among ternaries, and Sr$_{0.5}$Ce$_{0.5}$MnO$_3$ and Sr$_{0.5}$Ce$_{0.5}$CrO$_3$ among quaternaries to be promising candidates, as indicated by the overlap of calculated E[V$_{\text{O}}$] with the optimal range of 3.2-4.1~eV. Our findings highlight a promising chemical strategy for designing advaned STWS materials that enable efficient green H$_2$ generation and will be useful in the exploration of other higher-component oxide perovskites for STWS and beyond.

\section*{Conflicts of interest}
The authors declare no conflicts of interest.

\section*{Acknowledgements}
G.S.G. acknowledges financial support from Tata Consultancy Services (TCS). The authors acknowledge the computational resources provided by the Supercomputer Education and Research Centre (SERC), IISc and useful discussions with Dr. Sriram Goverapet Srinivasan of TCS. 

\section*{Data availability}
The data that support the findings of this study are available from the corresponding author upon reasonable request.

\bibliographystyle{unsrt}
\bibliography{rsc}

\begin{thebibliography}{10}

\bibitem{IEA2023}
{International Energy Agency}.
\newblock World energy outlook 2023.
\newblock Technical report, IEA, Paris, 2023.

\bibitem{Armaroli2007}
Nicola Armaroli and Vincenzo Balzani.
\newblock The future of energy supply: challenges and opportunities.
\newblock {\em Angew. Chem. Int. Ed.}, 46:52--66, 2007.

\bibitem{Turner1999}
John~A. Turner.
\newblock A realizable renewable energy future.
\newblock {\em Science}, 285(5428):687--689, 1999.

\bibitem{Falcone2021}
P.~M. Falcone, M.~Hiete, and A.~Sapio.
\newblock Hydrogen economy and sustainable development goals: review and policy
  insights.
\newblock {\em Curr. Opin. Green Sustain. Chem.}, 31:100506, 2021.

\bibitem{Steinfeld2005}
Aldo Steinfeld.
\newblock Solar thermochemical production of hydrogen—a review.
\newblock {\em Solar Energy}, 78(5):603--615, 2005.

\bibitem{Agrafiotis2015}
Christos Agrafiotis, Martin Roeb, and Christian Sattler.
\newblock A review on solar thermal syngas production via redox pair‐based
  water/carbon dioxide splitting thermochemical cycles.
\newblock {\em Renew. Sustain. Energy Rev.}, 42:254--285, 2015.

\bibitem{Chueh2009}
William~C. Chueh and Sossina~M. Haile.
\newblock Ceria as a thermochemical reaction medium for selectively generating
  syngas or methane from {H\(_2\)O} and {CO\(_2\)}.
\newblock {\em ChemSusChem}, 2(8):735--739, 2009.

\bibitem{muhich2017principles}
Christopher Muhich and Aldo Steinfeld.
\newblock Principles of doping ceria for the solar thermochemical redox
  splitting of h 2 o and co 2.
\newblock {\em J. Mater. Chem. A}, 5(30):15578--15590, 2017.

\bibitem{Bulfin2017}
Michael Bulfin, Johannes Vieten, Felix Call, Martin Lange, August Francke,
  Martin Roeb, and Christian Sattler.
\newblock Thermodynamics of ceria reduction and oxidation for solar
  thermochemical applications.
\newblock {\em J. Mater. Chem. A}, 5:1152--1160, 2017.

\bibitem{wexler2023materials}
Robert~B Wexler, Ellen~B Stechel, and Emily~A Carter.
\newblock Materials design directions for solar thermochemical water splitting.
\newblock {\em Solar Fuels}, pages 1--63, 2023.

\bibitem{McDaniel2013}
Anthony~H. McDaniel, Elizabeth~C. Miller, Dodi Arifin, Andrea Ambrosini,
  Eric~N. Coker, Ryan O'Hayre, William~C. Chueh, and Jijie Tong.
\newblock Sr$-$ and mn$-$doped laalo$_{3-\delta}$ for solar thermochemical
  h$_2$ and co production.
\newblock {\em Energy Environ. Sci.}, 6:2424--2428, 2013.

\bibitem{Vieten2016}
Johannes Vieten, Brendan Bulfin, Felix Call, Martin Lange, Marcus Schmücker,
  August Francke, Martin Roeb, and Christian Sattler.
\newblock Perovskite oxides for application in thermochemical air separation
  and oxygen storage.
\newblock {\em J. Mater. Chem. A}, 4(35):13652--13659, 2016.

\bibitem{witman2023defect}
Matthew~D Witman, Anuj Goyal, Tadashi Ogitsu, Anthony~H McDaniel, and Stephan
  Lany.
\newblock Defect graph neural networks for materials discovery in
  high-temperature clean-energy applications.
\newblock {\em Nat. Comp. Sci.}, 3(8):675--686, 2023.

\bibitem{naghavi2017giant}
S~Shahab Naghavi, Antoine~A Emery, Heine~A Hansen, Fei Zhou, Vidvuds Ozolins,
  and Chris Wolverton.
\newblock Giant onsite electronic entropy enhances the performance of ceria for
  water splitting.
\newblock {\em Nat. Comm.}, 8(1):285, 2017.

\bibitem{Gautam2020}
Gopalakrishnan Sai~Gautam, Ellen~B. Stechel, and Emily~A. Carter.
\newblock Exploring ca–ce–m–o (\(m\)=3d transition metal) oxide
  perovskites for solar thermochemical applications.
\newblock {\em Chem. Mater.}, 32(23):9964--9982, 2020.

\bibitem{wexler2023multiple}
Robert~B Wexler, Gopalakrishnan~Sai Gautam, Robert~T Bell, Sarah Shulda,
  Nicholas~A Strange, Jamie~A Trindell, Joshua~D Sugar, Eli Nygren, Sami
  Sainio, Anthony~H McDaniel, et~al.
\newblock Multiple and nonlocal cation redox in ca--ce--ti--mn oxide
  perovskites for solar thermochemical applications.
\newblock {\em Energy Environ. Sci.}, 16(6):2550--2560, 2023.

\bibitem{gao2023remarkable}
Ke~Gao, Xianglei Liu, Qi~Wang, Zhixing Jiang, Cheng Tian, Nan Sun, and Yimin
  Xuan.
\newblock Remarkable solar thermochemical co 2 splitting performances based on
  ce-and al-doped srmno 3 perovskites.
\newblock {\em Sustain. Energy Fuels}, 7(4):1027--1040, 2023.

\bibitem{heo2021double}
Su~Jeong Heo, Michael Sanders, Ryan O’Hayre, and Andriy Zakutayev.
\newblock Double-site substitution of ce into (ba, sr) mno3 perovskites for
  solar thermochemical hydrogen production.
\newblock {\em ACS Energy Lett.}, 6(9):3037--3043, 2021.

\bibitem{krzystowczyk2020substituted}
Emily Krzystowczyk, Xijun Wang, Jian Dou, Vasudev Haribal, and Fanxing Li.
\newblock Substituted srfeo 3 as robust oxygen sorbents for thermochemical air
  separation: correlating redox performance with compositional and structural
  properties.
\newblock {\em Phys. Chem. Chem. Phys.}, 22(16):8924--8932, 2020.

\bibitem{takeda1972synthesis}
Y~Takeda, T~Hashino, H~Miyamoto, F~Kanamaru, S~Kume, and M~Koizumi.
\newblock Synthesis of srnio3 and related compound, sr2ni2o5.
\newblock {\em J. Inorg. Nucl. Chem.}, 34(5):1599--1601, 1972.

\bibitem{sai2020first}
Gopalakrishnan Sai~Gautam, Ellen~B Stechel, and Emily~A Carter.
\newblock A first-principles-based sub-lattice formalism for predicting
  off-stoichiometry in materials for solar thermochemical applications: The
  example of ceria.
\newblock {\em Adv. Theory Simul.}, 3(9):2000112, 2020.

\bibitem{Kresse1996PRB}
Georg Kresse and Jürgen Furthmüller.
\newblock Efficient iterative schemes for \textit{ab initio} total‐energy
  calculations using a plane‐wave basis set.
\newblock {\em Phys. Rev. B}, 54(16):11169--11186, 1996.

\bibitem{kresse_ab_1993}
Georg Kresse and J.~Hafner.
\newblock \textit{Ab initio} molecular dynamics for liquid metals.
\newblock {\em Phys. Rev. B}, 47(1):558--561, 1993.

\bibitem{Kresse1999PRB_PAW}
Georg Kresse and David Joubert.
\newblock From ultrasoft pseudopotentials to the projector augmented‐wave
  method.
\newblock {\em Phys. Rev. B}, 59(3):1758--1775, 1999.

\bibitem{Sun2015PRL}
Jianwei Sun, Adrienn Ruzsinszky, and John~P. Perdew.
\newblock Strongly constrained and appropriately normed semilocal density
  functional.
\newblock {\em Phys. Rev. Lett.}, 115:036402, 2015.

\bibitem{sai2018evaluating}
Gopalakrishnan Sai~Gautam and Emily~A Carter.
\newblock Evaluating transition metal oxides within dft-scan and scan+ u
  frameworks for solar thermochemical applications.
\newblock {\em Phys. Rev. Mater.}, 2(9):095401, 2018.

\bibitem{long2020evaluating}
Olivia~Y Long, Gopalakrishnan Sai~Gautam, and Emily~A Carter.
\newblock Evaluating optimal u for 3 d transition-metal oxides within the scan+
  u framework.
\newblock {\em Phys. Rev. Mater.}, 4(4):045401, 2020.

\bibitem{swathilakshmi2023performance}
S~Swathilakshmi, Reshma Devi, and Gopalakrishnan Sai~Gautam.
\newblock Performance of the r2scan functional in transition metal oxides.
\newblock {\em J. Chem. Theory Comput.}, 19(13):4202--4215, 2023.

\bibitem{monkhorst1976special}
Hendrik~J Monkhorst and James~D Pack.
\newblock Special points for brillouin-zone integrations.
\newblock {\em Phys. Rev. B}, 13(12):5188, 1976.

\bibitem{Hellenbrandt2004CrystallogrRev}
Michael Hellenbrandt.
\newblock The inorganic crystal structure database (icsd)—present and future.
\newblock {\em Crystallogr. Rev.}, 10(1):17--22, 2004.

\bibitem{Ong2013ComputMaterSci}
Shyue~Ping Ong, William~D. Richards, Anubhav Jain, Geoffroy Hautier, Minseok
  Kocher, Shreyas Cholia, Daniel Gunter, Vincent~L. Chevrier, Kristin~A.
  Persson, and Gerbrand Ceder.
\newblock Python materials genomics ({pymatgen}): a robust, open‐source
  python library for materials analysis.
\newblock {\em Comput. Mater. Sci.}, 68:314--319, 2013.

\bibitem{wexler_factors_2021}
Robert~B. Wexler, Gopalakrishnan~Sai Gautam, Ellen~B. Stechel, and Emily~A.
  Carter.
\newblock Factors {Governing} {Oxygen} {Vacancy} {Formation} in {Oxide}
  {Perovskites}.
\newblock {\em J. Am. Chem. Soc.}, 143(33):13212--13227, August 2021.
\newblock Publisher: American Chemical Society.

\bibitem{Lan2003}
Y.~C. Lan, X.~L. Chen, and M.~He.
\newblock Structure, magnetic susceptibility and resistivity properties of
  {SrVO\(_3\)}.
\newblock {\em J. Alloys Compd.}, 354:95--98, 2003.

\bibitem{Zhu2012}
Zhong‐Li Zhu, Jian‐Hong Gu, Yu~Jia, and Xian Hu.
\newblock A comparative study of electronic structure and magnetic properties
  of {SrCrO\(_3\)} and {SrMoO\(_3\)}.
\newblock {\em Physica B Condens. Matter}, 407:1990--1994, 2012.

\bibitem{Wang2001}
Yong Wang, Jian Chen, and Xueyi Wu.
\newblock Preparation and gas‐sensing properties of perovskite‐type
  {SrFeO\(_3\)} oxide.
\newblock {\em Mater. Lett.}, 49(6):361--364, 2001.

\bibitem{Bezdicka1993}
Pavel Bezdicka, Alain Wattiaux, Jean‐Claude Grenier, Michel Pouchard, and
  Paul Hagenmuller.
\newblock Preparation and characterization of fully stoichiometric
  {SrCoO\(_3\)} by electrochemical oxidation.
\newblock {\em Z. Anorg. Allg. Chem.}, 619(1):7--12, 1993.

\bibitem{dong2023structure}
Fuxiao Dong, Gefei Lu, Qinghua Ma, Bojun Zhao, Haiou Wang, Dexuan Huo, and
  Weishi Tan.
\newblock Structure and magnetic properties of the manganite srmno3.
\newblock {\em J. Mater. Sci.: Mater. Electron.}, 34(36):2322, 2023.

\bibitem{barcellos2018}
Debora~R. Barcellos, Michael~D. Sanders, Jianhua Tong, Anthony~H. McDaniel, and
  Ryan~P. O’Hayre.
\newblock Bace$_{0.25}$mn$_{0.75}$o$_{3-\delta} -$ a promising perovskite-type
  oxide for solar thermochemical hydrogen production.
\newblock {\em Energy Environ. Sci.}, 11:3256--3265, 2018.

\end{thebibliography}

\newpage
\section{Electronic supplementary information}
\renewcommand{\theequation}{S\arabic{equation}}
\renewcommand{\thetable}{S\arabic{table}}
\renewcommand{\thefigure}{S\arabic{figure}}
\setcounter{equation}{0}
\setcounter{table}{0}
\setcounter{figure}{0}

\subsection*{SI. Overall solar thermochemical reaction}
\noindent The single-step thermal dissociation of \ce{H2O}:
\begin{equation} \label{eq:S1_H2O_direct}
\ce{H2O <=> H2 + 1/2 O2}
\end{equation}

\noindent The two-step solar thermochemical water splitting (STWS) cycle for a perovskite \ce{AMO3}, where M is a redox-active transition metal and A is a `large' cation:
\begin{equation} \label{eq:S2_STC_TR}
\ce{AMO3 -> AMO_{3-\delta} + \frac{\delta}{2} O2} \quad \text{(Thermal Reduction, TR)}
\end{equation}
\begin{equation} \label{eq:S3_STC_WS}
\ce{AMO_{3-\delta} + \delta H2O -> AMO3 + \delta H2} \quad \text{(Water Splitting, WS)}
\end{equation}
The overall reaction by summing (\ref{eq:S2_STC_TR}) and (\ref{eq:S3_STC_WS}):
\begin{equation} \label{eq:S4_STC_overall}
\ce{\delta H2O -> \delta H2 + \frac{\delta}{2} O2}
\end{equation}

\newpage
\subsection*{SII. Crystal Structures}
\noindent Table \ref{tab:icsd} lists the ternary \ce{SrMO3} polymorphs considered in this study, along with their space groups and collection codes from the inorganic crystal structure database (ICSD), where M~=~Ti, V, Cr, Mn, Fe, or Co. We used the listed structures as initial guesses for density functional theory (DFT) based calculations to identify the ground state polymorph of each SrMO$_3$ composition. 

\begin{table}[h!]
\centering
\caption{ICSD collection codes and space groups for ternary \ce{SrMO3} polymorphs.}
\label{tab:icsd}
\begin{tabular}{@{}lll@{}}
\toprule
System      & Space Group                                 & ICSD Collection Code \\
\midrule
\ce{SrTiO3} & I4/mcm                                      & 182762               \\
            & Pm$\bar{3}$m                                & 80873                \\
\addlinespace
\ce{SrVO3}  & Pm$\bar{3}$m                                & 88982                \\
\addlinespace
\ce{SrCrO3} & Pm$\bar{3}$m                                & 108903               \\
\addlinespace
\ce{SrMnO3} & C221                                        & 157936               \\
            & P6$_3$/mmc                                  & 185417               \\
            & Pm$\bar{3}$m                                & 17243                \\
            & P6$_3$                                      & 17245                \\
\addlinespace
\ce{SrFeO3} & Pm$\bar{3}$m                                & 92335                \\
\addlinespace
\ce{SrCoO3} & Pm$\bar{3}$m                                & 77142                \\
            & P12$_1$/m1                                  & 108896               \\
\bottomrule
\end{tabular}
\end{table}

\newpage
\subsection*{SIII. Convex hulls}
\begin{figure}[h!]
    \centering
    \includegraphics[width=\textwidth]{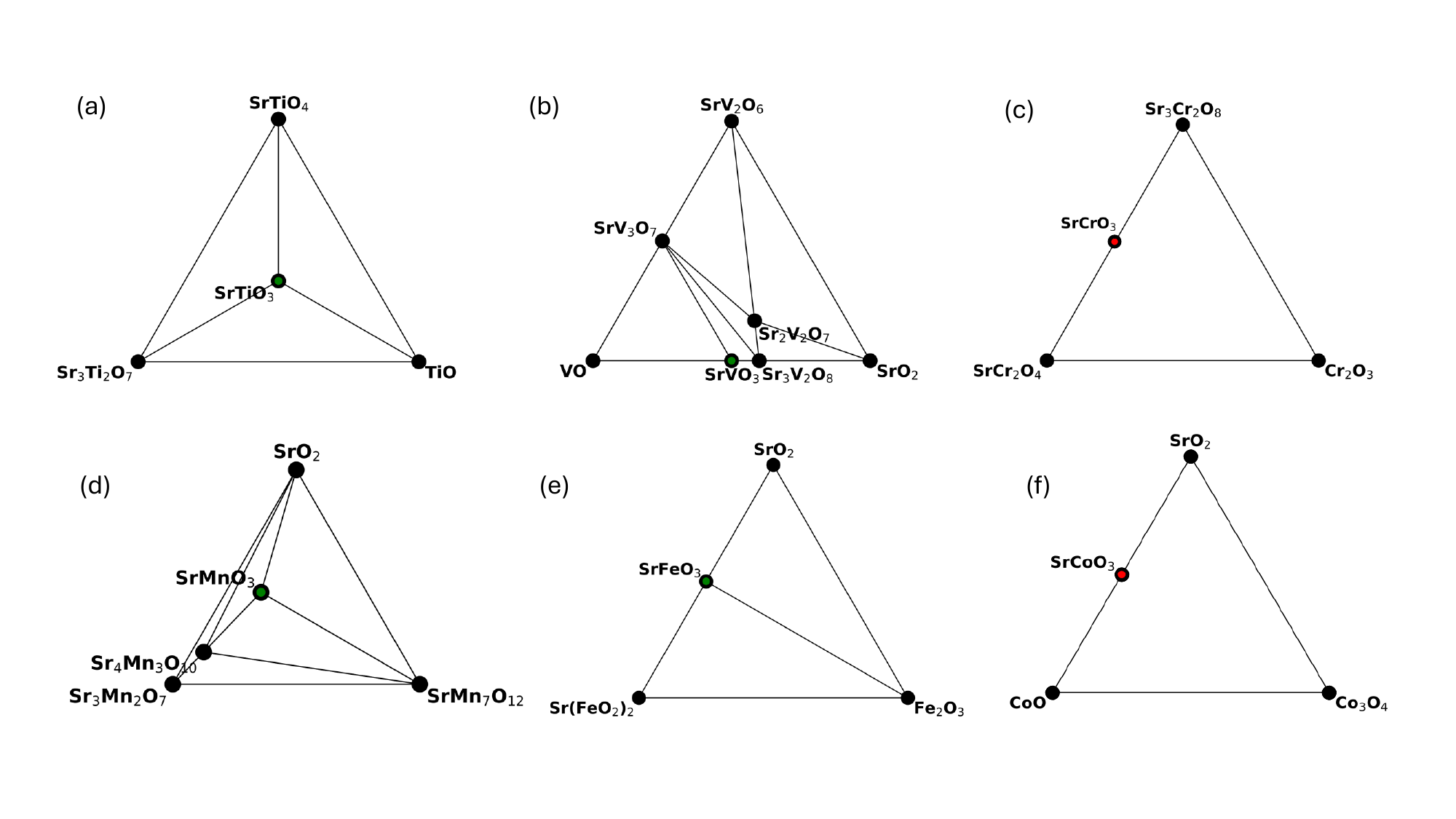}%
    \caption{Calculated 0~K convex hulls of ternary Sr-M-O systems, namely, (a) Sr-Ti-O, (b) Sr-V-O, (c) Sr-Cr-O, (d) Sr-Mn-O, (e) Sr-Fe-O, and (f) Sr-Co-O. Green (red) dots indicate that the perovskite compositions of interest lies on (above) the hull, with respect to the other competing phases (black dots). The thermodynamic decomposition products of SrCoO$_3$ are SrO$_2$, Sr$_2$Co$_2$O$_5$ and SrCo$_6$O$_{11}$, while that of SrCrO$_3$ are SrCr$_2$O$_4$ and Sr$_3$Cr$_2$O$_8$}.
    \label{fig:tcv}
\end{figure}

\begin{figure}[h!]
    \centering
    \includegraphics[width=\textwidth]{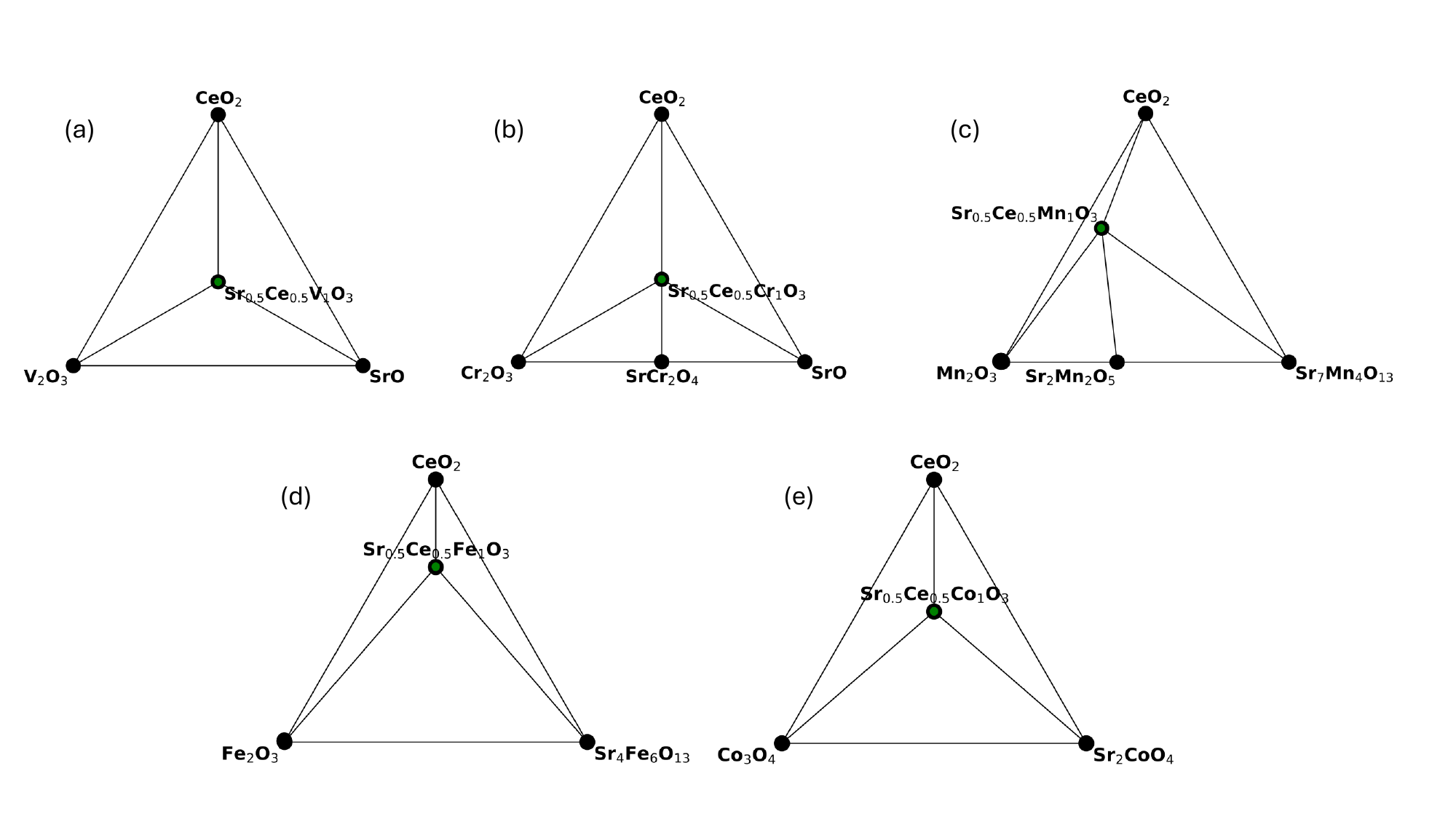}%
    \caption{Calculated 0~K convex hulls of quaternary Sr-Ce-M-O systems, namely, (a) Sr-Ce-V-O, (b) Sr-Ce-Cr-O, (c) Sr-Ce-Mn-O, (d) Sr-Ce-Fe-O, and (e) Sr-Ce-Co-O. Since quaternary convex hulls are 3D entities, we display relevant 2D projections containing the Sr$_{0.5}$Ce$_{0.5}$MO$_3$ perovskites of interest for ease of visualization. Notations used are similar to Figure~{\ref{fig:tcv}}.}
    \label{fig:qcv}
\end{figure}

\noindent We evaluated the thermodynamic stability of the ternary SrMO$_3$ and quaternary Sr$_{0.5}$Ce$_{0.5}$MO$_3$ perovskites by constructing the 0~K convex hull based on DFT-calculated total energies. Figure~\ref{fig:tcv} shows the ternary convex hulls for the Sr-M-O systems, while Figure~\ref{fig:qcv} shows relevant ternary projections of the quaternary convex hulls for the Sr-Ce-M-O systems. Perovskite compositions that lie on the hull (green dots) are predicted to be thermodynamically stable with respect to decomposition into other competing phases (black dots). Red dots indicate metastable/unstable perovskite compositions.

\clearpage
\subsection*{SIV. Oxidation states}
\begin{table}[h!]
\centering
\caption{Oxidation states of Sr and M cations in Sr-M-O ternaries in pristine and defect structures, based on calculated on-site magnetic moments. V exhibits mixed oxidation states in the pristine as well as defective SrVO$_3$ structure. Fe does not exhibit a noticeable change in its magnetic moment upon defect formation in SrFeO$_3$. Co exhibits both low-spin and high-spin configurations on its Co$^{4+}$ and Co$^{3+}$ sites upon defect formation in SrCoO$_3$.}
\label{tab:oxidation_states_srmo}
\begin{tabular}{|c|c|c|c|}
\hline
M-site & Compound & Sr oxidation state & M oxidation state \\
\hline
\multirow{2}{*}{Ti} & Pristine & Sr$^{2+}$ & Ti$^{4+}$ \\
& Defect   & Sr$^{2+}$ & Ti$^{4+/3+}$ \\
\hline
\multirow{2}{*}{V}  & Pristine & Sr$^{2+}$ & V$^{4+/3+}$ \\
& Defect   & Sr$^{2+}$ & V$^{4+/3+}$ \\
\hline
\multirow{2}{*}{Cr} & Pristine & Sr$^{2+}$ & Cr$^{4+}$ \\
& Defect   & Sr$^{2+}$ & Cr$^{4+/3+}$ \\
\hline
\multirow{2}{*}{Mn} & Pristine & Sr$^{2+}$ & Mn$^{4+}$ \\
& Defect   & Sr$^{2+}$ & Mn$^{4+/3+}$ \\
\hline
\multirow{2}{*}{Fe} & Pristine & Sr$^{2+}$ & Fe$^{4+}$ \\
& Defect   & Sr$^{2+}$ & Fe$^{4+}$ \\
\hline
\multirow{2}{*}{Co} & Pristine & Sr$^{2+}$ & Co$^{4+}$ \\
& Defect   & Sr$^{2+}$ & Co$^{4+/3+}$ \\
\hline
\end{tabular}
\end{table}

\begin{table}[h!]
\centering
\caption{Oxidation states of Ce and M cations in Sr$_{0.5}$Ce$_{0.5}$MO$_3$ perovskites in pristine and defect structures, based on calculated on-site magnetic moments. Sr exhibits the 2+ oxidation state in both the defective and pristine structures of all quaternaries considered. Although the Cr and Co systems contain Ce in its 4+ oxidation state, with potential for reduction to Ce$^{3+}$, we do not observe any noticeable change in the on-site magnetic moments of Ce$^{4+}$ upon defect formation in the quaternary structures, citing the sole participation of the transitiuon metal in the reduction. The Fe-containing quaternary exhibited an extent of disproportionation upon defect formation, i.e., Ce$^{3+}$ oxidized in addition to Fe$^{4+}$ reduction upon defect formation.}
\label{tab:oxidation_states}
\begin{tabular}{|c|c|c|c|}
\hline
M-site & Compound & Ce oxidation state & M oxidation state \\
\hline
\multirow{2}{*}{V}  & Pristine & Ce$^{3+}$ & V$^{4+/3+}$ \\
& Defect   & Ce$^{3+}$ & V$^{4+/3+}$ \\
\hline
\multirow{2}{*}{Cr} & Pristine & Ce$^{4+}$ & Cr$^{3+}$ \\
& Defect   & Ce$^{4+}$ & Cr$^{3+/2+}$ \\
\hline
\multirow{2}{*}{Mn} & Pristine & Ce$^{3+}$ & Mn$^{4+/3+}$ \\
& Defect   & Ce$^{3+}$ & Mn$^{4+/3+}$ \\
\hline
\multirow{2}{*}{Fe} & Pristine & Ce$^{3+}$ & Fe$^{4+/3+}$ \\
& Defect   & Ce$^{4+}$ & Fe$^{4+/3+}$ \\
\hline
\multirow{2}{*}{Co} & Pristine & Ce$^{4+}$ & Co$^{3+}$ \\
& Defect   & Ce$^{4+}$ & Co$^{3+/2+}$ \\
\hline
\end{tabular}
\end{table}
\clearpage

\subsection*{SV. Convex hull calculations}
\noindent ICSD collection codes of all ordered unary, binary, ternary, and quaternary Sr-Ce-M-O structures considered are compiled below, as grouped by M (transition metal, M = Ti, V, Cr, Mn, Fe, Co, or Ni), Ce, and Sr. Note that we did not consider intermetallic phases (e.g., Sr-M, Ce-M, M-M', Sr-Ce-M, etc.) in our convex hull calculations.

\setlength\LTleft{0pt}
\setlength\LTright{0pt}
\setlength{\LTpre}{0pt}
\setlength{\LTpost}{6pt}
\small
\setlength{\tabcolsep}{6pt}
\renewcommand{\arraystretch}{1.05}

\subsubsection*{Titanium-containing compounds (Ti)}
\begin{longtable}{@{}ll ll@{}}
\toprule
Compound & ICSD ID & Compound & ICSD ID \\
\midrule
\endfirsthead
\multicolumn{4}{c}{\tablename~\thetable\ (continued)}\\
\toprule
Compound & ICSD ID & Compound & ICSD ID \\
\midrule
\endhead
\midrule
\multicolumn{4}{r}{\emph{continued on next page}}\\
\endfoot
\bottomrule
\endlastfoot
\ce{Ti}          & 43416  & \ce{TiO}         & 15327 \\
\ce{TiO2}        & 9191   & \ce{Ti2O}        & 23574 \\
\ce{Ti2O3}       & 6095   & \ce{Ti3O}        & 20041 \\
\ce{Ti3O5}       & 75194  & \ce{Ti4O5}       & 77697 \\
\ce{Ti4O7}       & 6098   & \ce{Ti5O9}       & 9038 \\
\ce{Ti6O}        & 17009  & \ce{Ti6O11}      & 35121 \\
\ce{Ti7O13}      & 35122  & \ce{Ti8O15}      & 35123 \\
\ce{Ti9O17}      & 35124  & \ce{SrTi11O20}   & 71299 \\
\ce{Sr2Ti6O13}   & 10455  & \ce{Sr3Ti2O7}    & 63704 \\
\ce{SrTiO3}      & 182762 & \ce{SrTiO4}      & 194713 \\

\end{longtable}

\subsubsection*{Vanadium-containing compounds (\ce{V})}
\begin{longtable}{@{}ll ll@{}}
\toprule
Compound & ICSD ID & Compound & ICSD ID \\
\midrule
\endfirsthead
\multicolumn{4}{c}{\tablename~\thetable\ (continued)}\\
\toprule
Compound & ICSD ID & Compound & ICSD ID \\
\midrule
\endhead
\midrule
\multicolumn{4}{r}{\emph{continued on next page}}\\
\endfoot
\bottomrule
\endlastfoot
\ce{V}         & 43420  & \ce{VO}        & 28681  \\
\ce{VO2}       & 15889  & \ce{V2O3}      & 6286   \\
\ce{V2O5}      & 24042  & \ce{V3O5}      & 16445  \\
\ce{V3O7}      & 2338   & \ce{V4O7}      & 2775   \\
\ce{V4O9}      & 15041  & \ce{V5O9}      & 6097   \\
\ce{V6O11}     & 196    & \ce{V6O13}     & 15028  \\
\ce{V7O3}      & 77706  & \ce{V7O13}     & 197    \\
\ce{V8O}       & 166600 & \ce{V8O15}     & 424885 \\
\ce{V9O17}     & 424886 & \ce{V13O16}    & 77708  \\
\ce{V16O3}     & 77707  & \ce{SrV3O7}    & 72604  \\
\ce{SrV6O15}   & 98574  & \ce{SrV2O6}    & 65928  \\
\ce{Sr2V2O7}   & 10330  & \ce{Sr4V3O10}  & 73698  \\
\ce{Sr2VO4}    & 69000  & \ce{SrV6O11}   & 71868  \\
\ce{Sr3V2O8}   & 54655  & \ce{SrV4O9}    & 90926  \\
\ce{SrV13O18}  & 97949  & \ce{Sr3V2O7}   & 71320  \\
\ce{SrV10O15}  & 15273  &                 &        \\
\end{longtable}
\clearpage

\subsubsection*{Chromium-containing compounds (\ce{Cr})}
\begin{longtable}{@{}ll ll@{}}
\toprule
Compound & ICSD ID & Compound & ICSD ID \\
\midrule
\endfirsthead
\multicolumn{4}{c}{\tablename~\thetable\ (continued)}\\
\toprule
Compound & ICSD ID & Compound & ICSD ID \\
\midrule
\endhead
\midrule
\multicolumn{4}{r}{\emph{continued on next page}}\\
\endfoot
\bottomrule
\endlastfoot
\ce{Cr}        & 44731  & \ce{CrO}        & 109296 \\
\ce{CrO2}      & 9423   & \ce{CrO3}       & 16031  \\
\ce{Cr2O3}     & 25781  & \ce{Cr3O}       & 15904  \\
\ce{Cr5O12}    & 24299  & \ce{Cr8O21}     & 71297  \\
\ce{Sr9Cr5O18} & 91252  & \ce{SrCr2O4}    & 6132   \\
\ce{SrCr10O15} & 85053  & \ce{Sr2CrO4}    & 26944  \\
\ce{Sr4Cr3O9}  & 85045  & \ce{SrCrO3}     & 245834 \\
\ce{SrCrO4}    & 259674 & \ce{SrCrO4}     & 259675 \\
\ce{Sr4Cr3O10} & 257906 & \ce{Sr3Cr2O8}   & 85055  \\
\ce{SrCr2O7}   & 28411  &                  &        \\
\end{longtable}

\subsubsection*{Manganese-containing compounds (\ce{Mn})}
\begin{longtable}{@{}ll ll@{}}
\toprule
Compound & ICSD ID & Compound & ICSD ID \\
\midrule
\endfirsthead
\multicolumn{4}{c}{\tablename~\thetable\ (continued)}\\
\toprule
Compound & ICSD ID & Compound & ICSD ID \\
\midrule
\endhead
\midrule
\multicolumn{4}{r}{\emph{continued on next page}}\\
\endfoot
\bottomrule
\endlastfoot
\ce{Mn}        & 5248   & \ce{MnO}        & 9864   \\
\ce{MnO2}      & 393    & \ce{Mn2O3}      & 24342  \\
\ce{Mn2O7}     & 60821  & \ce{Mn3O4}      & 8355   \\
\ce{Mn5O8}     & 16956  & \ce{Sr7Mn4O15}  & 72332  \\
\ce{Sr3Mn2O7}  & 51215  & \ce{Sr5Mn5O13}  & 159660 \\
\ce{Sr2Mn2O5}  & 90184  & \ce{Sr2Mn2O5}   & 90183  \\
\ce{Sr2MnO4}   & 26560  & \ce{Sr7Mn4O13}  & 160291 \\
\ce{SrMnO3}    & 17243  &  \ce{SrMn7O12}   & 195757  \\
\ce{Sr4Mn3O10}    & 81351  &    &   \\
\end{longtable}

\subsubsection*{Iron-containing compounds (\ce{Fe})}
\begin{longtable}{@{}ll ll@{}}
\toprule
Compound & ICSD ID & Compound & ICSD ID \\
\midrule
\endfirsthead
\multicolumn{4}{c}{\tablename~\thetable\ (continued)}\\
\toprule
Compound & ICSD ID & Compound & ICSD ID \\
\midrule
\endhead
\midrule
\multicolumn{4}{r}{\emph{continued on next page}}\\
\endfoot
\bottomrule
\endlastfoot
\ce{Fe}          & 14754   & \ce{FeO}         & 31081 \\
\ce{Fe2O3}       & 7797    & \ce{Fe3O4}       & 5247 \\
\ce{Fe4O5}       & 434152  & \ce{Fe5O7}       & 430563 \\
\ce{Fe7O9}       & 430601  & \ce{Fe7O10}      & 135154 \\
\ce{Fe13O19}     & 238770  & \ce{Fe25O32}     & 430562 \\
\ce{SrFeO2}      & 173434  & \ce{SrFeO3}      & 58460 \\
\ce{Sr2FeO3}     & 251018  & \ce{Sr2FeO4}     & 74419 \\
\ce{Sr(FeO2)2}   & 94350   & \ce{Sr3(FeO3)2}  & 74434 \\
\ce{Sr2Fe2O5}    & 51318   & \ce{Sr2Fe2O5}    & 247823 \\
\ce{Sr3Fe2O5}    & 261512  & \ce{Sr3Fe2O7}    & 2648 \\
\ce{Sr4Fe4O11}   & 91064   & \ce{Sr4Fe6O12}   & 290200 \\
\ce{Sr4Fe6O13}   & 63621   & \ce{Sr8Fe8O23}   & 91063 \\
\ce{Sr25Fe30O77} & 118097  & \ce{Sr2(FeO2)3}  & 290200 \\
\end{longtable}
\clearpage

\subsubsection*{Cobalt-containing compounds (\ce{Co})}
\begin{longtable}{@{}ll ll@{}}
\toprule
Compound & ICSD ID & Compound & ICSD ID \\
\midrule
\endfirsthead
\multicolumn{4}{c}{\tablename~\thetable\ (continued)}\\
\toprule
Compound & ICSD ID & Compound & ICSD ID \\
\midrule
\endhead
\midrule
\multicolumn{4}{r}{\emph{continued on next page}}\\
\endfoot
\bottomrule
\endlastfoot
\ce{Co}          & 36675  & \ce{CoO}         & 9865 \\
\ce{CoO2}        & 88722  & \ce{Co3O4}       & 24210 \\
\ce{Sr2Co2O5}    & 162239 & \ce{Sr2Co2O5}    & 162240 \\
\ce{Sr2Co2O5}    & 173696 & \ce{Sr8Co8O23}   & 245305 \\
\ce{Sr6(CoO3)5}  & 173698 & \ce{Sr6(CoO3)5}  & 81312 \\
\ce{SrCoO}       & 155223 & \ce{SrCoO3}      & 108896 \\
\ce{Sr2CoO3}     & 99894  & \ce{Sr5(CoO3)4}  & 88628 \\
\ce{Sr3(CoO3)2}  & 182288 & \ce{SrCo6O11}    & 152279 \\
\ce{Sr2CoO4}     & 246483 &                  &        \\

\end{longtable}

\subsubsection*{Cerium-containing compounds (\ce{Ce})}
\begin{longtable}{@{}ll ll@{}}
\toprule
Compound & ICSD ID & Compound & ICSD ID \\
\midrule
\endfirsthead
\multicolumn{4}{c}{\tablename~\thetable\ (continued)}\\
\toprule
Compound & ICSD ID & Compound & ICSD ID \\
\midrule
\endhead
\midrule
\multicolumn{4}{r}{\emph{continued on next page}}\\
\endfoot
\bottomrule
\endlastfoot
\multicolumn{4}{c}{} \\
\ce{Ce}        & 41823   & \ce{CeO}       & 52886  \\
\ce{CeO2}      & 70257   & \ce{Ce2O3}     & 621706 \\
\ce{Ce7O12}    & 88754   & \ce{SrCeO3}    & 71352  \\
\ce{CeVO3}     & 99835   & \ce{CeCrO3}    & 28931  \\
\ce{CeMn7O12}  & 8773    & \ce{CeCrO3}    & 28931  \\
\ce{CeVO3}     & 99835   &                 &        \\
\end{longtable}

\subsubsection*{Strontium-containing compounds (\ce{Sr})}
\begin{longtable}{@{}ll ll@{}}
\toprule
Compound & ICSD ID & Compound & ICSD ID \\
\midrule
\endfirsthead
\multicolumn{4}{c}{\tablename~\thetable\ (continued)}\\
\toprule
Compound & ICSD ID & Compound & ICSD ID \\
\midrule
\endhead
\midrule
\multicolumn{4}{r}{\emph{continued on next page}}\\
\endfoot
\bottomrule
\endlastfoot
\multicolumn{4}{c}{} \\
\ce{Sr}    & 44721  & \ce{SrO}   & 19987 \\
\ce{SrO2}  & 24249  &            &       \\
\end{longtable}

\end{document}